\documentclass[%
 reprint,
 amsmath,amssymb, 
 aps,
prl,
]{revtex4-1}

\usepackage{graphicx}% Include figure files
\usepackage{dcolumn}% Align table columns on decimal point
\usepackage{bm}% bold math
\usepackage{amsmath}
\usepackage{amsfonts}
\usepackage{amssymb}
\usepackage{epsfig}

\usepackage[pdftex]{color}
\usepackage[latin1]{inputenc}
\usepackage[T1]{fontenc}
\usepackage{lmodern}
\usepackage{bbm,array,theorem,enumerate,stmaryrd}
\usepackage{fancyhdr}
\usepackage{amsmath,amsxtra,amscd,amssymb,latexsym,comment,lineno,rotating}
\usepackage{textcomp}
\usepackage[toc,page]{appendix}
\usepackage{etoolbox}
\usepackage{setspace}
\usepackage{subfigure}
\usepackage{stfloats}

\begin{document}

\preprint{APS/123-QED}

\title{Localization of ions within one-, two- and three-dimensional Coulomb crystals\\ by a standing wave optical potential}

\author{Thomas Laupr\^etre$^1$, Rasmus B. Linnet$^1$, Ian D. Leroux$^{2}$, Aur\'elien Dantan$^1$}

\author{Michael Drewsen$^1$}%
 \email{drewsen@phys.au.dk}
\affiliation{%
 $^1$Department of Physics and Astronomy, Aarhus University, DK-8000 Aarhus C, Denmark \\
$^2$National Research Council Canada, Ottawa, Ontario, Canada K1A 0R6
}%

\date{\today}

\begin{abstract}
We demonstrate light-induced localization of Coulomb-interacting particles in multi-dimensional structures. Subwavelength localization of ions within small multi-dimensional Coulomb crystals by an intracavity optical standing wave field is evidenced by measuring the difference in scattering inside symmetrically red- and blue-detuned optical lattices and is observed even for ions undergoing substantial radial micromotion.
These results are promising steps towards the structural control of ion Coulomb crystals by optical fields as well as for complex many-body simulations with ion crystals or for the investigation of heat transfer at the nanoscale, and have potential applications for ion-based cavity quantum electrodynamics, cavity optomechanics and ultracold ion chemistry.
\end{abstract}


\maketitle

The interplay between Coulomb-interacting particles and optical potentials provides an interesting setting for quantum simulations with ions~\cite{Blatt2012,Richerme2014,Jurcevic2014} and the emulation of various fundamental many-body models~\cite{Schneider2012,Bermudez2012}, such as the Frenkel-Kontorova model of friction~\cite{Garcia2007,Benassi2011,Pruttivarasin2011,Cetina2013,Bylinskii2015,Fogarty2015,Gangloff2015,Bylinskii2016}, Ising spin-models~\cite{Porras2004,Britton2012} or generalized Dicke models~\cite{Genway2014}.
It also potentially allows the investigation of ion dynamics in quantum potentials~\cite{Bushev2004,Cormick2012}, the control of the crystalline structure of large Coulomb crystals~\cite{Horak2012}, or the study of energy transport in coupled oscillators systems~\cite{Pruttivarasin2011,Ramm2014,Ruiz2014,Freitas2015,Abdelrahman2016}.
Moreover, the application of cavity-generated optical potentials to trapped ions has natural applications for the enhancement of the ion-light coupling in cavity QED experiments~\cite{Herskind2009Realization,Albert2011,Begley2016}, the implementation of cavity-cooling schemes~\cite{Leibrandt2009,Karpa2013,Fogarty2016}, or the investigation of nanofriction with dynamically deformable substrates~\cite{Fogarty2015}.
Optical forces on ions can also be potentially exploited in ``hybrid'' settings, {\it e.g.} for coupling atoms with nanomechanical resonators~\cite{Camerer2011,Genes2011} or in cold chemistry experiments involving ions and neutrals~\cite{Grier2009,Zipkes2010,Schmid2010,Cetina2012,Huber2014}.

Until now, experimental investigations of ion dynamics in optical lattices have been limited to single ions~\cite{Katori1997,Enderlein2012,Linnet2012,Schmiegelow2016} or small one-dimensional crystals~\cite{Karpa2013,Bylinskii2015,Begley2016} in radiofrequency traps.
Following an earlier observation of anomalous ion diffusion in a standing wave field~\cite{Katori1997}, subwavelength localization of single ions in optical lattices has recently been demonstrated in various settings~\cite{Enderlein2012,Linnet2012,Karpa2013} and, very recently, the role of the optical-lattice-induced friction in small ion strings has been investigated~\cite{Bylinskii2015,Gangloff2015,Bylinskii2016}.
Extending these studies to higher-dimensional crystals is a prerequisite for the use of ion crystals in more complex many-body investigations.
However, in radiofrequency traps, rf-induced micromotion is known to critically affect the dynamics of single ions in optical potentials~\cite{Schneider2010,Enderlein2012,Linnet2012}.
In the case of two- or three-dimensional crystals, ions away from the rf-field-free nodal line possess micromotion-induced radial kinetic energy which can easily exceed the thermal energy of the ion inside the lattice by orders of magnitude \cite{Berkeland1998,Landa2012}.
Since the longitudinal and transverse motional degrees of freedom are coupled in such crystals -  either intrinsically due to the Coulomb interaction or extrinsically because of asymmetries or imperfections of the trapping potentials - assessing the feasibility of optical confinement under realistic conditions is thus critical.

In this Letter, we report on the simultaneous subwavelength localization of up to eight $^{40}$Ca$^+$ ions in one-, two- and three-dimensional Coulomb crystals by the application of an intracavity optical standing wave field.
The pinning of the ions along the direction of the optical lattice potential is inferred by measuring the scattering induced by intracavity fields having equal intensity, but being symmetrically red- or blue-detuned with respect to the 3d$^2$D$_{3/2}\rightarrow$ 4p$^2$P$_{1/2}$ transition.
The results are in agreement with a simple scattering model and demonstrate that the light-induced localization of multi-ion crystals is not impeded by the crystal structural dimension and micromotion, even when the kinetic energy of the radial rf-induced motion is much larger (by a factor up to $\sim30$) than the motional energy of the ion inside the lattice.
As such, they represent an important step towards the implementation of complex many-body models with ions or the control of the crystalline structure of large Coulomb crystals.

\begin{figure}%
\includegraphics[width=\columnwidth]{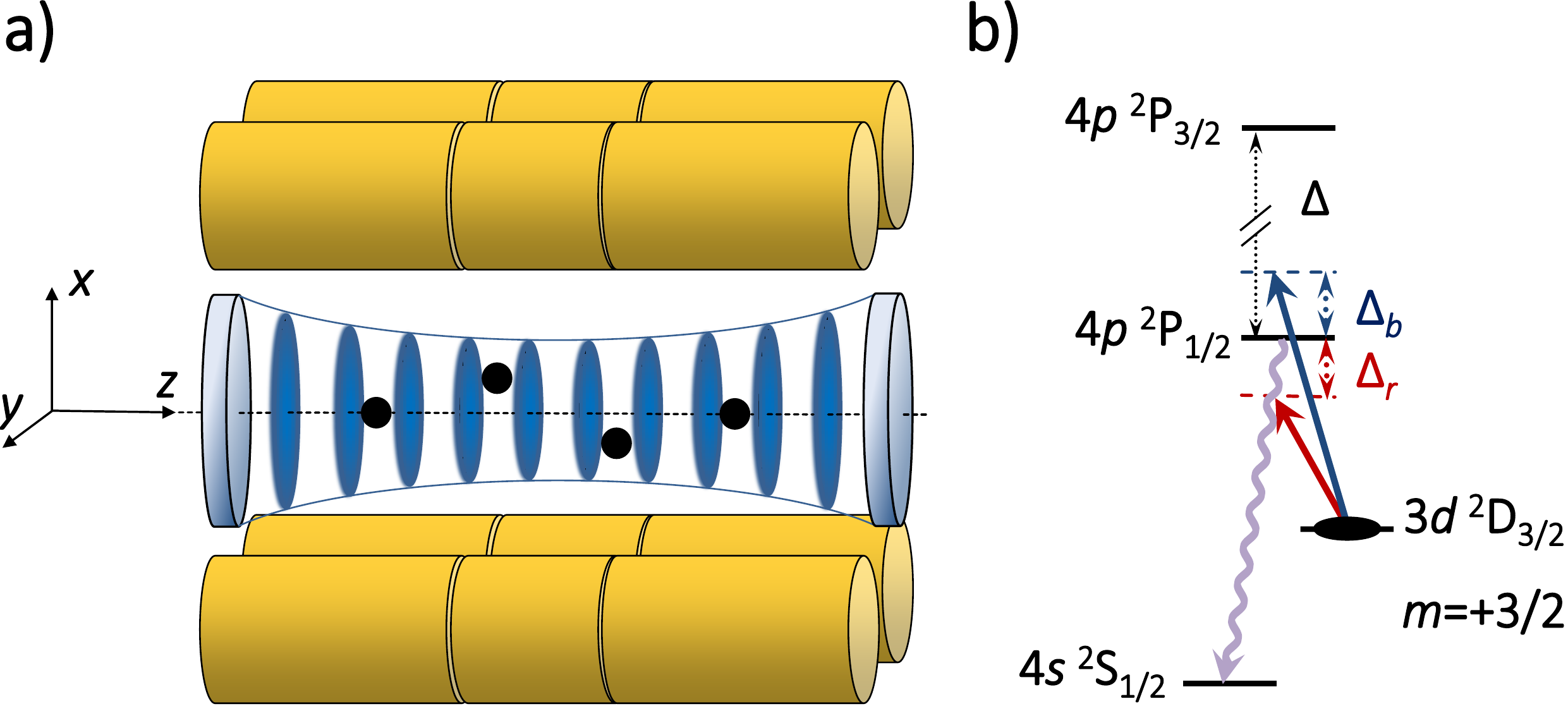}%
\caption{(color online).
(a) Experimental setup: ions trapped in a linear Paul trap and arranged in a Coulomb crystal are pinned along the $z$-direction by an intracavity standing wave field.
(b) Diagram of the relevant energy levels in $^{40}$Ca$^+$. The ions, initially pumped into the $m=+3/2$ Zeeman sublevel of the D$_{3/2}$ state after Doppler cooling, are pinned by a $\sigma^-$-circularly polarized intracavity field, which is either red- or blue-detuned with respect to the D$_{3/2}\rightarrow\mathrm{P}_{1/2}$ transition at 866~nm by $\Delta_{b,r}/(2\pi)\sim\pm0.76~$THz (red and blue arrows). The detuning between the P$_{1/2}$ and P$_{3/2}$ states is $\Delta/(2\pi)\sim6.7~$THz. The lattice-induced excitation of an ion to the P$_{1/2}$ state is detected by collecting the 397~nm photon emitted when the ion decays to the S$_{1/2}$ state.}%
\label{fig1}%
\end{figure}

A number $N$ of $^{40}$Ca$^+$ ions is produced and confined in a linear Paul trap described in detail in~\cite{Herskind2008,Herskind2009Positionning}  (Fig.~\ref{fig1}).
The trap operates at an rf-frequency of $\sim3.98$ MHz, with axial and radial trap frequencies in the range 70-110 kHz and 180-400 kHz, respectively.
By the combined application of light fields close to resonance with the S$_{1/2}\rightarrow$ P$_{1/2}$ and D$_{3/2}\rightarrow$ P$_{1/2}$ transitions, and in the presence of a 2.2 G bias magnetic field along the $z$ axis (see Fig.~\ref{fig1}), ions are first Doppler-cooled for $62\;\mu\mathrm{s}$ then optically pumped for $75\;\mu\mathrm{s}$ to the $|\mathrm{D}_{3/2},m=+3/2\rangle$ state (>98\% efficiency per ion on average).
The resulting Coulomb crystal has a typical inter-ion distance of the order of $\sim20$ $\mu$m.
A 11.8 mm-long linear Fabry-Perot cavity with moderate finesse ($\sim3000$) and waist radius $\sim37$ $\mu$m at 866~nm allows for the generation of a standing wave along the $z$ axis with intensity up to $\sim500$ kW/cm$^2$ at the center of the trap.
Ions were previously positioned at the absolute center of the optical cavity following the method of~\cite{Linnet2014}.
After switching off the cooling and optical pumping fields, a $\sigma^-$-circularly polarized standing wave field detuned either to the blue or the red side of the D$_{3/2}\rightarrow$ P$_{1/2}$ transition by $\pm 0.76\;\mathrm{THz}$ is ramped up adiabatically for $2\;\mu\mathrm{s}$ and held at its maximum level for $1\;\mu\mathrm{s}$.
An independent and absolute calibration of the lattice potential depth experienced by a single ion as a function of the intensity transmitted out of the cavity is used as a reference, and a maximum lattice depth of $\sim25$ mK corresponding to a lattice vibrational frequency of~$\sim3.7$ MHz \cite{suppl} can be reached at this detuning in the limit of the available laser power.
When an ion is excited to the P$_{1/2}$ state by the intracavity field, it leaves the $|\mathrm{D}_{3/2},m=+3/2\rangle$ state with 97\% probability by subsequently decaying to either the $|\mathrm{D}_{3/2},m=\pm1/2\rangle$ states  (3\%) or predominantly to the S$_{1/2}$ state (94\%) where it no longer interacts with the standing wave.
The 397 nm photons scattered in the latter case are detected by an intensified CCD camera with a detection efficiency of $\sim1.7\times 10^{-4}$.
The probability to interact with the lattice after a scattering event and subsequent heating of the ion is very low, even allowing for the 2\% of cases where it decays to the $|\mathrm{D}_{3/2},m=+1/2\rangle$, whose coupling to the intracavity field is three times weaker due to the smaller Clebsch-Gordan coefficient.

\begin{figure}
\includegraphics[width=\columnwidth]{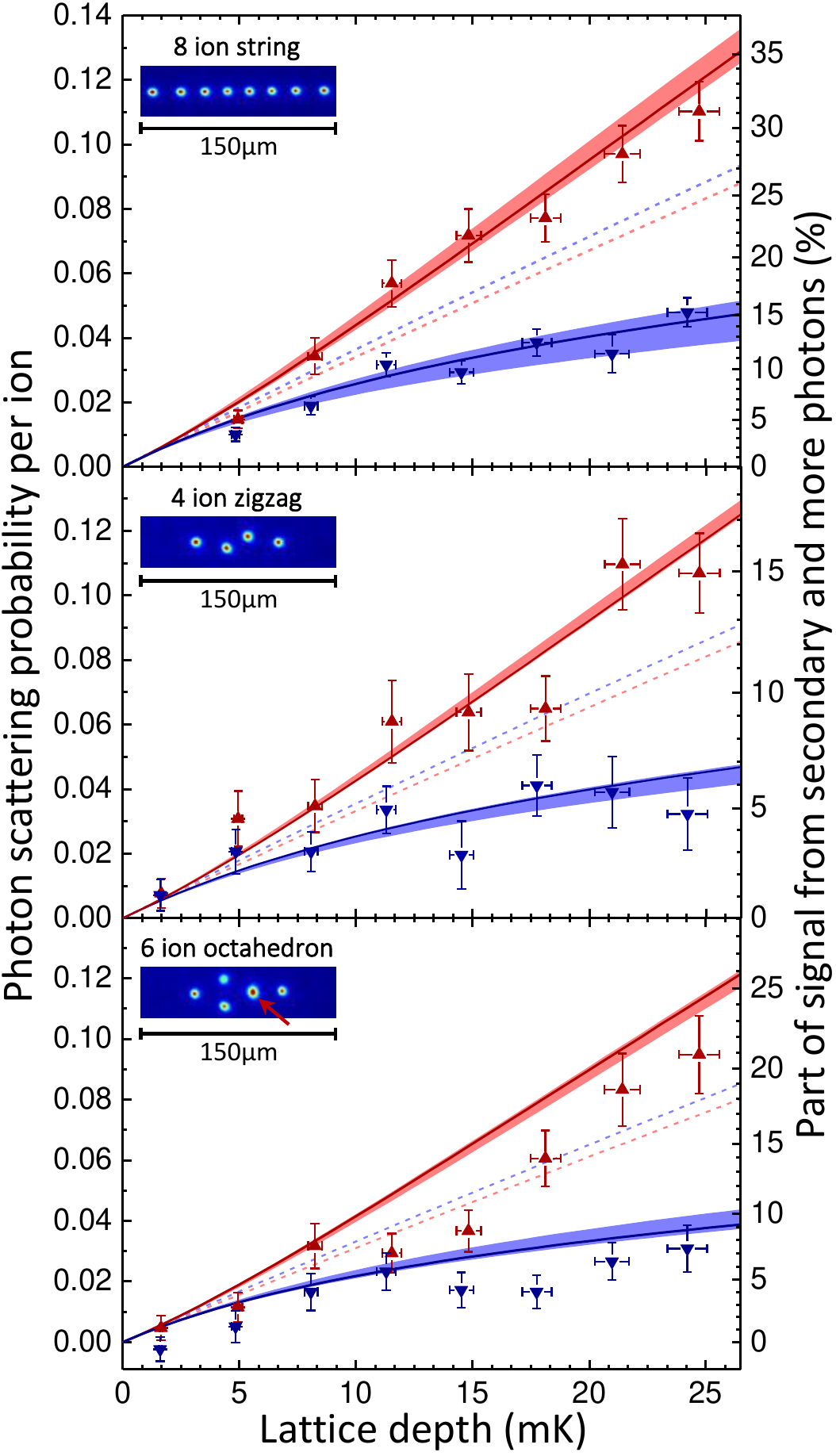}
\caption{(color online).
Photon scattering probability per ion as a function of the optical lattice depth for an 8-ion string (top), a two-dimensional 4-ion zigzag crystal (center) and a three-dimensional 6-ion octahedron crystal (bottom).
The red up-triangles/blue down-triangles are experimental data points for the red-/blue-detuned lattices. Each data point corresponds to the repetition of approximately $2\times 10^6$ sequences.
The red and blue shaded areas are the theoretical scattering probabilities from the single ion model of Ref.~\cite{Linnet2012} with the initial temperature obtained from the ion pictures and its error bar represented by the thickness.
The red and blue continuous lines are the results from the fits with the same model for which the initial temperature is left as a free parameter.
The red and blue dashed lines show the theoretical scattering probabilities expected for delocalized ions. The right scale represents the fraction of the signal due to secondary and later emitted photons when considering binomial statistics for the scattering per ion. The insets on the top-left corners are fluorescence images of the ion crystals. For the bottom picture, the red arrow head indicates the spot at which the fluorescence of the two out-of-plane ions is overlapped.
}
\label{fig2}
\end{figure}

The measured scattering probability per ion as a function of the optical lattice depth is plotted in Fig.~\ref{fig2} for three different spatial configurations of the ions: a one-dimensional 8-ion string, a two-dimensional 4-ion zigzag crystal and a three-dimensional 6-ion octahedron crystal. The insets in Fig.~\ref{fig2} show fluorescence images of these crystals, for which the axial and radial trap frequencies are (71,350), (87,185) and (105,192) kHz, respectively. For the zigzag and octahedron crystals, a small voltage ($\sim100$~mV) was applied to one pair of diagonal electrodes in order to induce a slight asymmetry in the radial trapping potentials and improve the long term stability of the structure.
The octahedron crystal has a structure that has an S$_4$ symmetry, {\it i.e.} it maps back onto itself by a combination of a 90$^{\circ}$ rotation around the axis connecting the two ions furthest apart and a reflection across the central plane perpendicular to this axis.
Configurational changes of the crystal in presence of the highest intensity of the red-detuned standing wave field (lattice depth $\sim25$ mK) typically occurred at a rate of $\sim2$ s$^{-1}$ for the zigzag chain, and $\sim4$ s$^{-1}$ in case of the octahedron crystal.
As detailed in the Supplemental Material \cite{suppl}, the initial average axial temperature of the ions in each configuration is evaluated by measuring their position variance based on the detected images of their fluorescence prior to the application of the optical lattice, and calculating numerically the frequency of the normal modes of motion from the axial and radial frequencies of the trap~\cite{Knunz2012,Rajagopal2016}. For the three crystals shown in Fig.~\ref{fig2}, the axial temperatures are found to be $3.6\pm 1.1$, $3.5\pm 0.5$, and $3.1\pm 0.5$ mK, respectively.

\begin{figure*}
\centering
\includegraphics[width=\textwidth]{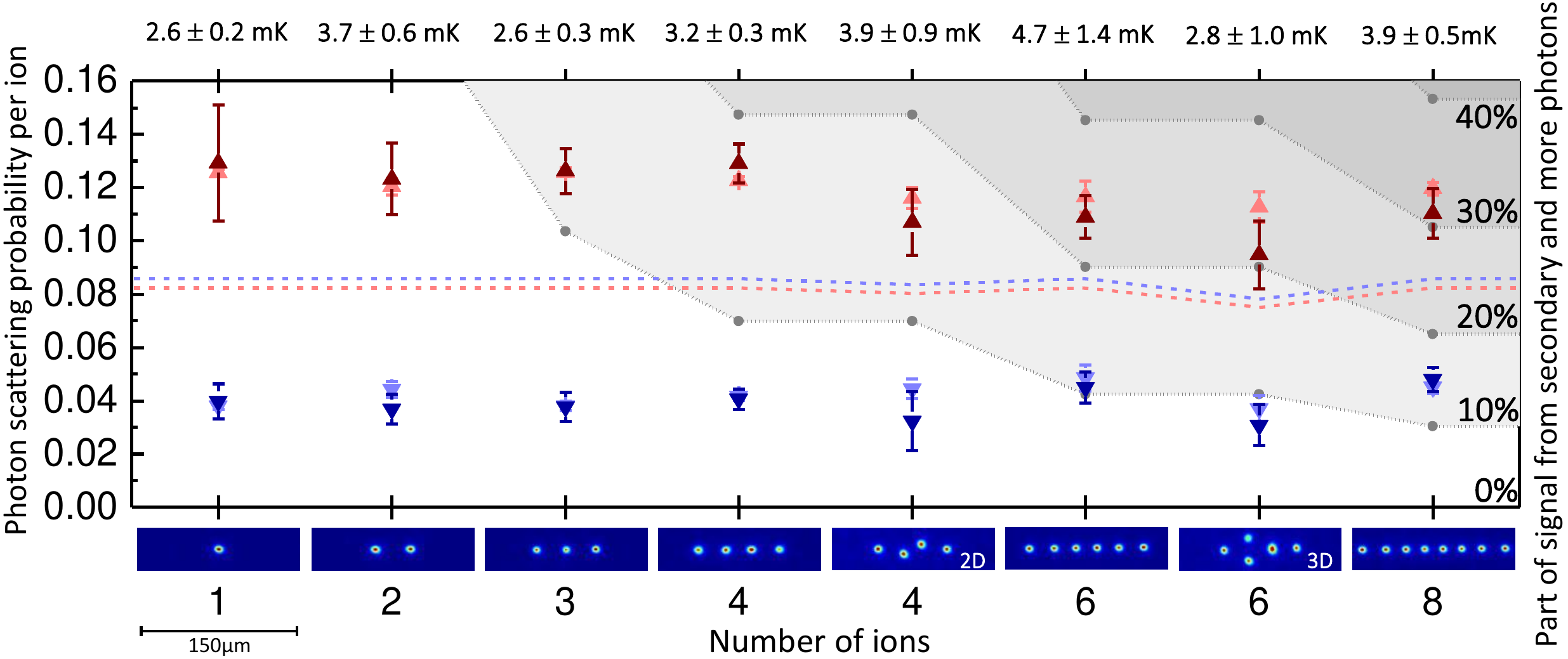}%
\caption{(color online).
Photon scattering probability per ion for $\sim25\;\mathrm{mK}$-deep lattices as a function of the number of ions in the crystal. The red up-triangles/blue down-triangles are experimental data points for the red/blue detuned lattices.
The light-red and light-blue symbols show the corresponding results obtained from fitting to the theoretical model, while the red and blue dashed lines show the predicted theoretical results for delocalized ions.
The different shaded gray areas indicate the fraction of signal due to secondary and later scattering events.}
\label{fig3}
\end{figure*}

In these conditions the initial position distribution of each ion extends over several lattice periods, so that the variation of the background trapping potential over one lattice period is small with respect to the initial thermal energy. As the lattice potential is adiabatically raised, each ion thus experiences approximately the same potential. Moreover, given the relatively large inter-ion distances used in this work, the lattice-induced forces quickly overcome the trapping and Coulomb forces along the axis. As such, the ions can be expected to independently localize close to the minima (maxima) of intensity of the blue (red)-detuned standing wave field, as demonstrated with a single ion in~\cite{Linnet2012}.
We therefore base the analysis on the single-ion model of Ref.~\cite{Linnet2012} which, given the temperature of the initial thermal position distribution of ions, determines the final position distribution in the lattice after adiabatic ramp-up and yields a prediction for the scattering probability per ion.

The red and blue shaded areas in Fig.~\ref{fig2} show the model predictions computed with the independently measured intial temperature for each configuration. These curves show a good agreement with the experimental data points (up and down triangles). As a reference, the red and blue dashed lines show the theoretical expectations for delocalized ions. 
Even though the lattice depths and detunings from the P$_{1/2}$ state are equal, the slight asymmetry in the scattering probabilities is due to the non-negligible excitation to the P$_{3/2}$ state, which is taken into account in the model.
For the zigzag and the octahedron crystals, the lower intensity experienced by ions off the trap axis and not at the center of the intracavity field beam has also been accounted for. 
Because the octahedron crystal is not as stable as the string or the zigzag crystals, heating of the entire structure (followed by slow recrystallization) occurs more frequently over the long acquisition period needed to accumulate photon counts.
Therefore, a reduced measured scattering probability might be expected for this configuration.

The solid lines on Fig~\ref{fig2} show fits to the theoretical model with the initial temperature left as a free parameter. The initial temperatures extracted from the fits are $3.9\pm 0.5$, $3.9\pm 0.9$ and $2.8\pm 1.0$ mK, respectively, in very good agreement with the temperatures independently determined from the fluorescence images. 
Defining the average probability to be pinned in a single lattice site by the fraction of the energy distribution lying below the lattice depth, the inferred energy distributions show that ions are pinned in a single well of the deepest blue-detuned lattice field ($\sim24$ mK) with probabilities 96\%, 95\% and 97\%, respectively, clearly indicating that subwavelength localization in the optical potential is achieved.  

In principle, the dynamics of a multi-ion crystal are more complex than for a single ion, since, as soon as one ion in the crystal scatters a photon from the standing wave field excitation, it will decay with almost unit probability to a state which is not affected by the lattice, thereby changing the total potential energy of the system.
The change in potential energy of the ions still affected by the optical lattice could possibly lead to a change in the localization process as their configuration changes, which would then alter subsequent scattering events.
This experiment should, however, be insensitive to such effects for two reasons.
First, the interaction time with the lattice is never much longer than the period of the highest-frequency normal mode.
Consequently, the system has too little time to relax and redistribute thermal energy between ions after a scattering event.
Second, since the initial position distribution of each ion extends over several lattice sites, crystal distorsions due to pinning or depinning alter the potential energy of the system by only a fraction of the initial thermal energy.
Another possible complexity is imperfect optical pumping, which leads to an exponential decrease with $N$ of the probability to have initially all ions in the responsive state.
However, the effect is less critical than that of a sudden depinning since the lattice is raised adiabatically in presence of the unaffected ions, and it only results in reduced scattering probability.

In any case, in the limit where the probability of exciting more than one ion in each experimental sequence is low, the detected scattering is essentially that of the first ion to be excited, thus providing information on its average position distribution inside the lattice potential before any depinning event. We show on the right scale of the graphs of Fig.~\ref{fig2} the fraction of signal that is due to secondary or later scattering events as a function of the scattering probability per ion, when considering binomial statistics~\cite{suppl}. While this fraction is less than 15\% for all blue-detuned lattice depths, the model is still observed to give good predictions for the data taken with red-detuned lattices for which this fraction may be substantially higher. The single-ion model can therefore be considered to give accurate predictions in our experimental conditions.

Last, and most importantly, micromotion could impede the light-induced localization, as it is known to cause off-resonant excitation of the ions when the lattice potential is raised~\cite{Enderlein2012,Linnet2012}. For the 8-ion string of Fig.~\ref{fig2}, the kinetic energy associated with the residual excess micromotion is not expected to be larger than 0.4~mK for the most external ions~\cite{suppl}. But the cases of two- and three-dimensional crystals are more critical, as the axial and transverse motional degrees of freedom are coupled. Even in a perfectly quadrupolar potential, ions located away from the rf-field free axis experience a driven micromotion whose kinetic energy is typically much larger than the axial thermal energy or the depth of the optical potential. For the crystals of Fig.~\ref{fig2}, we evaluate the micromotion radial kinetic energy of the off-axis ions to be in the worse case of the octahedron crystal $\sim800$ mK~\cite{suppl}. We nevertheless observe in Fig.~\ref{fig2} a clear difference in the scattering propabilities in the blue- and red-detuned lattices for all configurations.

To confirm the one-dimensional localization by the optical lattice potential, similar experiments were performed for crystals with various number of ions and structural configurations. Figure~\ref{fig3} shows the scattering probability per ion for both symmetrically blue- and red-detuned lattices with a fixed depth of $\sim25\;\mathrm{mK}$, as a function of the number of ions. The indicated initial temperatures are determined by fitting the average scattering probability from the standing wave field with the single ion theoretical model.

We have demonstrated subwavelength localization of ions in multi-dimensional ion Coulomb crystals using an intracavity optical standing wave field. We note that 
the interaction time with the standing wave field was purposely kept short in these experiments in order to minimize multi-ion excitation and check the agreement with the theoretical expectations. However, this time could be extended further by either increasing the standing wave field detuning, cooling the ions to lower temperatures or by operating on the S$_{1/2}\rightarrow$ P$_{1/2}$ or S$_{1/2}\rightarrow$ P$_{3/2}$ transitions.
The fact that rf-induced micromotion in these multidimensional ion crystal structures does not impede the lattice-induced localization is very promising for achieving deterministic control of the crystalline structure of cold charged plasmas, as predicted \textit{e.g.} by numerical simulations~\cite{Horak2012}. The application of such \textit{intracavity} optical fields could also be used for exploring the complex dynamics of Coulomb particles in cavity-generated potentials~\cite{Cormick2012,Fogarty2015}, for cavity optomechanics experiments with ion crystals~\cite{Fogarty2016} and for enhancing the ion-photon coupling for efficient ion crystal-based photon memories and counters~\cite{Zangenberg2012,Clausen2013}.

We acknowledge financial support from the European Commission (STREP PICC and ITN CCQED), the Carlsberg Foundation, the Villum Foundation and the Sapere Aude Initiative. We are grateful to Haggai Landa for providing the software for calculating the ion crystals' normal modes.

%\nocite{*} %Even non-cited BibTeX-Entries will be shown.
%\bibliographystyle{apalike} %Style of Bibliography: plain / apalike / alpha/ amsalpha / ...
\bibliographystyle{apsrev}
\bibliography{Bibliography} %You need a file 'literature.bib' for this.

\begin{widetext}
\pagebreak

\part{Supplemental Material\newline}

\section{Lattice parameters}

The lattice potential along the longitudinal trap axis is $U(z)=U_0\sin^2(kz)$, with $k$ the optical field wavevector. The lattice depth, in units of temperature, is
\begin{equation*}
T_{\textrm{latt}}=\frac{\vert U_0\vert}{k_\textrm{B}},
\end{equation*}
where $k_\textrm{B}$ is the Boltzmann constant. The vibrational frequency (in Hz) of a particle of mass $M$ at the bottom of a well of the lattice is
\[
\nu_{\textrm{latt}}=\frac{k}{2\pi}\sqrt{\frac{2k_\textrm{B}T_{\textrm{latt}}}{M}}.
\]

%%%%%%%%%%%%%%%%%%%%%%%%%%%%%%%%%%%
%%%%%%%%%%%%%%%%%%%%%%%%%%%%%%%%%%
\section{Photon scattering probability}

We assume binomial statistics for the scattering of the ions in an $N$-ion crystal with a scattering probability per ion $p$. The probability for the ions in the crystal to scatter $N_p$ photon is given by 
\[
P_N(N_p)=\left(^N_{N_p}\right)p^{N_p}(1-p)^{N-N_p},
\]
so the average number of scattered photons is $N\times p$. The probability to scatter no photon at all is then $P_N(0)=(1-p)^{N}$. The complementary of $P_N(0)$ is the probability to scatter at least 1 photon: $P_N(N_p\geq1)=1-(1-p)^{N}$. As the first photon emitted exists as soon as some photon has been emitted, the average number of first emitted photons is $1\times P_N(N_p\geq1)=1-(1-p)^{N}$. The proportion $f$ of signal due to secondary or later emitted photons is finally given by
\begin{equation}
f=1-\frac{1-(1-p)^{N}}{Np}.
\end{equation}

%%%%%%%%%%%%%%%%%%%%%%%%%%%%%%%%%%%
%%%%%%%%%%%%%%%%%%%%%%%%%%%%%%%%%%
\section{Micromotion kinetic energy}

For any ion located at a distance $r_0$ from the rf-field free point in a ``non peculiar'' crystal, i.e. a crystal where the pseudo-potential limit correctly predicts the mean positions of the ions, the amplitude of excess micromotion $A_{\mu,u}$ along any direction $u$ is identical to the single ion case and is given as a function of the parameter $q_u$ of the trap along the considered direction by~\cite{Landa2012}:
\[
A_{\mu,u}=r_0\frac{q_u}{2}.
\]
The associated average kinetic energy for a particule of mass $M$ is then given by
\[
E_{\mu,u}^{\textrm{kin}}=\frac{1}{4}M\Omega_{rf}^2A_{\mu,u}^2\equiv \frac{1}{2}k_{\textrm{B}}T_{\mu,u}^{\textrm{kin}},
\]
for which we define here a temperature associated to the driven micromotion.

%%%%%%%%%%%%%%%%%%%%%%%%%
%%%%%%%%%%%%%%%%%%%%%
\section{Determination of the initial temperature}

The initial temperature of the ions is estimated from fluorescence pictures taken at the end of the Doppler cooling part of the sequence. We follow the method developped in~\cite{Norton2011,Knunz2012} for single ions and for strings of ions in~\cite{Rajagopal2016} by introducing the crystal's normal modes of motion~\cite{James1998} . We have extended this method to two- and three-dimensional crystals, which we detail below before tackling the experimental procedure.

\subsection{Theory}

\subsubsection{Normal mode decomposition}

We consider an $N$-ion crystal in the pseudo-potential consisting of an axial harmonic trap with angular frequency $\omega_z$ and a radial harmonic trap with angular frequency $\omega_r$. The temperature is assumed to be low enough that the motional amplitude of each ion is small compared to the average distance between ions. A first order Taylor expansion of the total potential can be written as the sum of the potential of uncoupled harmonic oscillators, which define the $3N$ normal modes associated with the three-dimensional motion.

The excursions of the $m$-th ion from its equilibrium position $(x_m^0,y_m^0,z_m^0)$ are denoted by $(\delta x_m,\delta y_m,\delta z_m)$, so that, in a given direction $u$, $u_m(t)=u_m^0+\delta u_m(t)$. Let $(b_l^p)_{1\leq l\leq3N}$ and $\lambda_p$ be the $3N$ coordinates and the eigenvalue of the $p$-th eigenvector of the matrix describing the approximated potential in the Lagrangian normalized by $\omega_z^2$. The amplitude $Q_p$ of the $p$-th mode with frequency $\omega_p=\omega_z\sqrt{\lambda_p}$ can then be written as
\[
Q_p=\sum_{m=1}^Nb_m^p\delta x_m+\sum_{m=1}^Nb_{N+m}^p\delta y_m+\sum_{m=1}^Nb_{2N+m}^p\delta z_m,
\]
With these notations we have in each direction
\begin{equation*}
\delta x_m =\sum_{p=1}^{3N}b_m^pQ_p,\hspace{0.3cm}
\delta y_m =\sum_{p=1}^{3N}b_{N+m}^pQ_p,\hspace{0.3cm}
\delta z_m =\sum_{p=1}^{3N}b_{2N+m}^pQ_p.
\end{equation*}

Because these normal modes are uncoupled, one has:
\begin{equation*}
\langle \delta x_m^2\rangle =\sum_{p=1}^{3N}(b_m^p)^2\langle Q_p^2\rangle,\hspace{0.3cm}
\langle \delta y_m^2\rangle =\sum_{p=1}^{3N}(b_{N+m}^p)^2\langle Q_p^2\rangle,\hspace{0.3cm}
\langle \delta z_m^2\rangle =\sum_{p=1}^{3N}(b_{2N+m}^p)^2\langle Q_p^2\rangle.
\end{equation*}
and,
\begin{equation*}
\langle \dot{\delta x_m}^2\rangle =\sum_{p=1}^{3N}(b_m^p)^2\langle\dot{Q}_p^2\rangle,\hspace{0.3cm}
\langle \dot{\delta y_m}^2\rangle =\sum_{p=1}^{3N}(b_{N+m}^p)^2\langle\dot{Q}_p^2\rangle,\hspace{0.3cm}
\langle \dot{\delta z_m}^2\rangle =\sum_{p=1}^{3N}(b_{2N+m}^p)^2\langle\dot{Q}_p^2\rangle.
\end{equation*}
In the most general case,  let $T_{m,u}$ be the $m$-th ion's temperature in the direction $u$ ($u=x,y,z$) and $T_p$ the temperature associated with the $p$-th normal mode. The preceding relations imply that
\begin{equation*}
T_{m,x} =\sum_{p=1}^{3N}(b_m^p)^2T_p,\hspace{0.3cm}
T_{m,y} =\sum_{p=1}^{3N}(b_{N+m}^p)^2T_p,\hspace{0.3cm}
T_{m,z} =\sum_{p=1}^{3N}(b_{2N+m}^p)^2T_p.
\end{equation*}
In general, the normal modes do not necessarily have the same temperature. As each mode is a harmonic oscillator, we have though the relation
\[
k_{\textrm{B}}T_p=M\langle\dot{Q}_p^2\rangle=M\omega_p^2\langle Q_p^2\rangle,
\]
and the variances of the spatial excursions in the three directions can be expressed as a function of the normal mode temperatures as
\begin{equation*}
\langle \delta x_m^2\rangle =\frac{k_{\textrm{B}}}{M}\sum_{p=1}^{3N}(b_{m}^p)^2\frac{T_p}{\omega_p^2},\hspace{0.3cm}
\langle \delta y_m^2\rangle =\frac{k_{\textrm{B}}}{M}\sum_{p=1}^{3N}(b_{N+m}^p)^2\frac{T_p}{\omega_p^2},\hspace{0.3cm}
\langle \delta z_m^2\rangle =\frac{k_{\textrm{B}}}{M}\sum_{p=1}^{3N}(b_{2N+m}^p)^2\frac{T_p}{\omega_p^2}.
\end{equation*}

\subsubsection{Thermal equilibrium}

We now assume complete thermal equilibrium of the system, {\it i.e.} that all the normal modes have the same temperature $T$. Because $\sum_{p=1}^{3N}(b_i^p)^2=1$ $\forall i$, this implies that all ions also have the same temperature $T$, identical in each direction. The variance of the spatial excursions become
\begin{equation*}
\langle \delta x_m^2\rangle =\frac{k_{\textrm{B}}T}{M}\sum_{p=1}^{3N}\frac{(b_{m}^p)^2}{\omega_p^2},\hspace{0.3cm}
\langle \delta y_m^2\rangle =\frac{k_{\textrm{B}}T}{M}\sum_{p=1}^{3N}\frac{(b_{N+m}^p)^2}{\omega_p^2},\hspace{0.3cm}
\langle \delta z_m^2\rangle =\frac{k_{\textrm{B}}T}{M}\sum_{p=1}^{3N}\frac{(b_{2N+m}^p)^2}{\omega_p^2}.
\end{equation*}
Let us define constants associated with each ion and direction
\begin{equation*}
\gamma_{m,x}^2 =\sum_{p=1}^{3N}\frac{(b_{m}^p)^2}{\lambda_p},\hspace{0.3cm}
\gamma_{m,y}^2 =\sum_{p=1}^{3N}\frac{(b_{N+m}^p)^2}{\lambda_p},\hspace{0.3cm}
\gamma_{m,z}^2 =\sum_{p=1}^{3N}\frac{(b_{2N+m}^p)^2}{\lambda_p},
\end{equation*}
so that one can write in the usual harmonic oscillator form
\begin{equation*}
\langle \delta x_m^2\rangle =\frac{k_{\textrm{B}}T}{M\omega_z^2}\gamma_{m,x}^2,\hspace{0.3cm}
\langle \delta y_m^2\rangle =\frac{k_{\textrm{B}}T}{M\omega_z^2}\gamma_{m,y}^2,\hspace{0.3cm}
\langle \delta z_m^2\rangle =\frac{k_{\textrm{B}}T}{M\omega_z^2}\gamma_{m,z}^2.
\end{equation*}
It is clear that ions in a string at a temperature $T$ have a lower position excursion than the single ion at the same temperature because $\gamma_{m,z}<1$ $\forall m$. Note that all variances have been expressed as a function of the frequency of the center of mass mode frequency in the axial direction, because the potential was initially normalized to it.

\subsection{Experimental analysis}

\subsubsection{Evaluating the $\gamma$ parameters}

\noindent $\bullet$ To obtain the values of the $\gamma$ parameters for a given configuration, {\it i.e.} for a certain number of ions and fixed axial and radial frequencies, we make use of a Matlab program provided by Haggai Landa to numerically calculate from the pseudo-potential the mode coordinates $(b_l^p)_{1\leq l\leq3N}$ and their eigenvalues $\lambda_p$. Given the experimental uncertainty on the trap frequencies, a minimization of the difference of the calculated ion positions with the measured positions on fluorescence pictures is performed as a function of the trap frequencies to obtain more precise values.
\\

\noindent $\bullet$ In the radial direction, the modes coordinates have to be projected to account for the fact that the image plane is at $45^{\circ}$ from the radial trap axes. If $r_m$ is the $m$-th ion's coordinate in the radial direction in the image plane, one has
\begin{align*}
r_m &=\frac{x_m+y_m}{\sqrt{2}},\\
\delta r_m &=\sum_{p=1}^{3N}\frac{b_m^p+b_{N+m}^p}{\sqrt{2}}Q_p=\sum_{p=1}^{3N}b_{m,rad}^pQ_p\\
\langle \delta r_m^2\rangle &=\frac{k_{\textrm{B}}T}{M\omega_z^2}\gamma_{m,rad}^2
\end{align*}
where the parameter $\gamma_{m,rad}$ is defined as
\begin{equation*}
\gamma_{m,rad}^2 =\sum_{p=1}^{3N}\frac{(b_{m,rad}^p)^2}{\lambda_p}=\sum_{p=1}^{3N}\frac{(b_m^p+b_{N+m}^p)^2}{2\lambda_p} \\
\end{equation*}

\subsubsection{Picture analysis}

\noindent $\bullet$ The thermal distribution of a harmonic oscillator in presence of a damping force and Brownian motion, as it is the case in the final stage of cooling, is almost gaussian~\cite{Blatt1986}. All modes have thus a gaussian distribution, which implies that the distribution of each excursion $\delta u_m$ is also gaussian.
\\

\noindent $\bullet$ The fluorescence spot observed on a picture is in each direction the convolution of the distribution of $\delta u_m$ with the Point Spread Function of the imaging system, which we assume to be gaussian with variance $\sigma_{\textrm{res}}^2$. The final recorded spot is thus gaussian with a variance in each direction given by
\[
\sigma_{m,u}^2=\langle \delta u_m^2\rangle+\sigma_{\textrm{res}}^2=\frac{k_{\textrm{B}}T}{M\omega_z^2}\gamma_{m,u}^2+\sigma_{\textrm{res}}^2.
\]
The resolution of our imaging system is $\sigma_{\textrm{res,ax}}^2=2.23\pm0.02~\mu$m in the axial direction and $\sigma_{\textrm{res,rad}}^2=2.09\pm0.02~\mu$m in the radial direction, and the pixellisation of our images with the experimental magnification corresponds to $0.92~\mu$m/pixel.
\\

\noindent $\bullet$ In the analyzed direction, the spot of the ion is integrated along the orthogonal direction and fitted by a gaussian function. The analysis is performed with Matlab with a Trust Region algorithm and the error bars are given by the 95\% confidence bounds.
\\

\noindent $\bullet$ In principle, the analysis should be carried out by processing all ions and directions at once with the equilibrium temperature $T$ as sole free parameter. In practice, we analyze all ions only in the $z$-direction for the following reasons:
\begin{itemize}
\item For strings, the lack of coupling between the radial and longitudinal degrees of freedom makes it possible to consider thermal equilibrium in the longitudinal direction separately.
\item For 2D and 3D configurations, because of the coupling between all degrees of freedom, the situation is more complex: ideally, without any breaking of degeneracy in the pseudo-potential, the zigzag and octahedron configurations tend to rotate around the axial trap axis. Experimentally, we obtain stable configurations by introducing a small bias voltage on one pair of diagonal electrodes in order to introduce an asymmetry between the two radial frequencies. The exact value of this asymmetry plays an important role when numerically calculating the radial spatial extensions. When the asymmetry is low, a mode with very low frequency corresponding to a large radial extension of the off-axis ions is present as a vestige of the degeneracy.
This mode tends to disappear with increasing asymmetry, thus reducing the spatial extension of ions in the radial direction. This phenomenon translates into relatively large variations of the $\gamma$ values for off-axis ions in the radial direction. As an example, for the 6 ion octahedron crystal, the superimposed off-axis ions can have their radial $\gamma$ value modified by 12\% when the relative difference of frequency is changed by 10\%. In contrast, the $\gamma$ values in the axial direction have variations of the order of $\sim10^{-4}$. We thus analyze ions in the axial direction only, and evaluate that the error in the temperature measurement due to the uncertainty on the $\gamma$ values should not exceed $\sim 10^{-3}$.
\\
\end{itemize}

\noindent $\bullet$ Typically, we observe that the application of this method leads to relative error bars on the initial temperature estimation of the order of 20\%. We discuss below potential sources of systematic errors whose effects are found to be within this experimental uncertainty.

\subsection{Systematic errors}

\subsubsection{Doppler cooling}

The damping force due to Doppler cooling could lead to frequency shifts of the normal modes. A very conservative estimate based on the detunings and Rabi frequencies of the cooling lasers gives a systematic relative error lower than $6\cdot10^{-3}$ on the temperature. 

\subsubsection{Effect of micromotion}

In each direction, the secular motion of the $m$-th ion is written as a function of the normal mode amplitudes
\[
u_{m,\textrm{sec}}(t)=u_m^0+\sum_{p=1}^{3N}b_{m_u+m}^pQ_p(t)=u_m^0+\delta u_{m,\textrm{sec}}(t),
\] 
where $m_u=0,N,2N$ depending on the direction.

According to \cite{Landa2012}, in an ideal trap and for a non peculiar crystal, the amplitude of the micromotion depends on the position $u$ and the $q_u$ parameter along the considered direction as $A_{\mu,u}=u\frac{q_u}{2}$. If the highest frequency of the secular motion is small compared to the trap rf frequency, the total trajectory of the $m$-th ion can be written at first order in $q_u$:
\[
u_m(t)=\left(u_m^0+\sum_{p=1}^{3N}b_{m_u+m}^pQ_p(t)\right)\left(1+\frac{q_u}{2}\cos(\Omega_{rf}t)\right)=\delta u_{m}(t)+u_m^0\left(1+\frac{q_u}{2}\cos(\Omega_{rf}t)\right).
\]

\noindent $\bullet$ The position distribution to consider is then modified compared to the ideal case as it is given by the distribution of $\delta u_{m}$ which includes the ordinary micromotion and the position distribution of the excess micromotion $u_m^0\frac{q_u}{2}\cos(\Omega_{rf}t)$.
\begin{itemize}
\item First, the part without excess micromotion $\delta u_{m}$ has a gaussian distribution and has its variance given by $\langle \delta u_m^2\rangle=\langle \delta u_{m,\textrm{sec}}^2\rangle\left(1+\frac{q^2}{8}\right)$.

In the case of a string along the trap axis, an experimental upper bound of $q_{z}$ has been determined to be $\sim 5\cdot 10^{-4}$ in the conditions of these experiments, which leads to a negligible change of the position distribution and a relative error of $\sim3\cdot10^{-8}$ on the temperature measurement.

In the case of 2D or 3D crystals, since $q_z$ is very small and because of the coupling between degrees of freedom, the value to consider along the axial direction is the effective parameter $q'_z\sim\left(\frac{q_{\textrm{rad}}}{4}\right)^2$ \cite{Landa2012}, whose value, for $q_{\textrm{rad}}\sim 0.14$, is $\sim 1.2\cdot 10^{-3}$ in our experimental conditions. This leads also in this case to a negligible change of the position distribution and a relative error of $\sim2\cdot10^{-7}$ on the temperature measurement.

\item Therefore, the relative broadening of the position distribution in presence of micromotion is for all presented cases dominated by the amplitude of the excess micromotion. For all our experiments, the excess axial micromotion amplitude does not exceed 20~nm for the most external ions, whereas the width of the spatial distribution is at least $\sim 1$ $\mu$m. A very conservative estimate would give a corresponding relative error on the temperature measurement of $\sim4\cdot10^{-2}$.
\end{itemize}
Since the excess micromotion amplitude is negligible in comparison with the spatial thermal distribution, the position distribution stays gaussian and one still has the relation:
\[
\langle (u_m-u_m^0)^2\rangle\approx\langle \delta u_m^2\rangle\approx\langle \delta u_{m,\textrm{sec}}^2\rangle=\sum_{p=1}^{3N}(b_{m_u+m}^p)^2\langle Q_p^2\rangle.
\]
Therefore, one still has in the thermal equilibrium hypothesis
\[
\langle (u_m-u_m^0)^2\rangle\approx\frac{k_{\textrm{B}}T}{M\omega_z^2}\gamma_{m,u}^{'2},
\]
with
\[
\gamma_{m,u}^{'2}=\sum_{p=1}^{3N}\frac{(b_{m}^p)^2}{\lambda'_p},
\]
where the value of $\gamma'_{m,u}$ now includes any shift of frequency of the normal modes due to micromotion in the eigenvalue $\lambda'_p=\omega_p^{'2}/\omega_z^2$.
\\

\noindent $\bullet$  In the case of a string along the trap axis, the frequencies of the axial normal modes are unchanged because there is no coupling between radial and axial directions. In the case of 2D or 3D crystals, the frequencies of the normal modes will in general be shifted, as compared to the pseudo-potential case, because of the coupling between degrees of freedom \cite{Landa2012}.
The exact values of the modes frequencies could in principle have quite an influence on the temperature results, and only a complete Floquet-Lyapunov computation would tell about the relative error made on the frequencies. Because we have low $q_{\textrm{rad}}$ values for our multi-dimensional crystals, we can expect the shift to be small \cite{Kaufmann2012}. As an example, if we consider a relative frequency shift of the order of $3\cdot10^{-2}$, such as observed in \cite{Kaufmann2012}, this would lead to a relative error on the temperature lower than $6\cdot10^{-2}$.
\\

\noindent $\bullet$ Last, let us point out that the temperature $T$ defined with this method corresponds to the thermal kinetic energy (i.e from the secular motion) and, as such, does not carry information about the total momentum distribution, which includes the driven micromotion. According to \cite{Blatt1986}, the position and velocity distributions of a harmonic oscillator in presence of micromotion as well as damping and random forces stay approximately gaussian. One can then write for each of the terms 
\begin{equation*}
u_p=b_{m_u+m}^pQ_p(t)\left(1+\frac{q_u}{2}\cos(\Omega_{rf}t)\right)
\end{equation*} the following relation
\begin{equation*}
\langle \dot{u}_p^2\rangle =\omega_p^{'2}\langle u_p^2\rangle\left(1+\frac{q_u^2\Omega_{rf}^2}{8\omega_p^{'2}}\right)\approx\omega_p^{'2}(b_{m_u+m}^{p})^2\langle Q_p^2\rangle\left(1+\frac{q_u^2\Omega_{rf}^2}{8\omega_p^{'2}}\right).
\end{equation*}
This gives an averaged kinetic energy along the considered direction $u$ equal to
\begin{equation*}
E_{u}^{\textrm{kin}} =\frac{M}{2}\sum_{p=1}^{3N}\omega_p^{'2}(b_{m_u+m}^{p})^2\langle Q_p^2\rangle\left(1+\frac{q_u^2\Omega_{rf}^2}{8\omega_p^{'2}}\right)+E_{\mu,u}^{\textrm{kin}}=\frac{k_{\textrm{B}}}{2}\sum_{p=1}^{3N}T_p(b_{m_u+m}^{p})^2\left(1+\frac{q_u^2\Omega_{rf}^2}{8\omega_p^{'2}}\right)+E_{\mu,u}^{\textrm{kin}},
\end{equation*}
which becomes, in case of thermal equilibrium,
\[
E_{u}^{\textrm{kin}}=\frac{k_{\textrm{B}}T}{2}\sum_{p=1}^{3N}(b_{m_u+m}^{p})^2\left(1+\frac{q_u^2\Omega_{rf}^2}{8\omega_p^{'2}}\right)+E_{\mu,u}^{\textrm{kin}}.
\]
For our experiments and along the trap axis, we would have at maximum $\frac{q_z^{'2}\Omega_{rf}^2}{8\omega_z^{2}}\sim\frac{\left(\frac{q_{\textrm{rad}}}{4}\right)^4\Omega_{rf}^2}{8\omega_z^{2}}\sim 10^{-4}$.
The momentum distribution along the trap axis excluding the excess micromotion can thus also be considered to be only thermal, i.e. given by the secular motion only, and directly observable from the position distribution.

Note for example that for a single ion on axis, the radial temperature extracted from its observed position distribution along the radial direction would give only half of its total kinetic energy, as half of it lies into the ordinary radial micromotion.

\end{widetext}

\end{document}